\documentclass[twocolumn,5p]{elsarticle}

\usepackage[version=4]{mhchem}
\usepackage{subcaption}
\usepackage{standalone}
\usepackage{hyperref}
\usepackage{siunitx}

\begin{document}

\begin{frontmatter}
\title{Modelling conduction cooling of superconducting accelerator magnets using a thermal thin shell approximation}

\author[ku,cern]{Emma Vancayseele\corref{cor1}}
\ead{emma.vancayseele@gmail.com}

\author[cern]{Erik Schnaubelt}
\ead{erik.schnaubelt@cern.ch}

\author[uliege]{Louis Denis}
\ead{louis.denis@uliege.be}

\author[uliege]{Christophe Geuzaine}
\ead{cgeuzaine@uliege.be}

\author[cern]{Arjan Verweij}
\ead{arjan.verweij@cern.ch}

\author[cern]{Mariusz Wozniak}
\ead{mariusz.wozniak@cern.ch}

\address[ku]{KU Leuven, Leuven, Belgium}
\address[cern]{CERN, Meyrin, Switzerland}
\address[uliege]{University of Liège, Liège, Belgium}
\cortext[cor1]{Corresponding author}

\begin{abstract}
Understanding the thermal behaviour of superconducting accelerator magnets is essential to ensure their stable and reliable operation. This work presents an extension of the Finite Element Quench Simulator (FiQuS) Multipole module to include collar and pole regions of accelerator magnets, which influences the overall thermal response. A thermal thin shell approximation (TSA), which is shown to be effective from previous works, is employed to model thermal insulation layers efficiently, replacing an insulation surface mesh. The main novelty of this work lies in the development of a method to model the thermal connection between the magnet winding and the collar and pole regions via the TSA. 

To assess the accuracy and computational efficiency of this method, temperature and field variations are computed for a current ramp-up scenario. The thermal solution is coupled to a fully resolved magnetodynamic solution to capture the interaction between thermal and electromagnetic behaviour. The results obtained with the TSA are then compared to classical finite element (FE) solutions with explicitly meshed insulation domains. The TSA predicts the maximum temperature within 2–4\,\% of the reference solution while substantially reducing mesh complexity and achieving up to a 5 times speed-up in computation time. While the TSA has traditionally been employed for short-duration quench simulations with high heat fluxes between magnet turns, these results demonstrate its reliability and efficiency for current ramp scenarios with low heat fluxes, significantly expanding its application range beyond what has been previously reported in the literature.

To illustrate potential applications of this new functionality, conduction cooling through the collar region is studied, comparing different cooling configurations and collar materials.

\end{abstract}


\end{frontmatter}

\section{Introduction}
The thermal stability of superconducting magnets is of critical importance for particle accelerators using such technology. At the European Laboratory for Particle Physics (CERN), the Simulation
of Transient Effects in Accelerator Magnets (STEAM)~\cite{Bortot2018, steam_website} framework provides a collection of tools to model the transient behaviour of these magnets. Within this framework, the Finite Element Quench Simulator (FiQuS)~\cite{Vitrano2023} is a coupled magnetodynamic-thermal finite-element based tool used to model the behaviour of a variety of magnets types (e.g. high-temperature superconducting pancake coils~\cite{Atalay2024}, multipole magnets~\cite{Vitrano2023}, canted cosine-theta magnets \cite{wozniak2024}). FiQuS relies on the Gmsh Python application programming interface~\cite{geuzaine2009gmsh} for geometry and mesh creation, and GetDP~\cite{getdp_1998, GetDP, schnaubelt_cerngetdp} for solving the governing equations.

Quench simulations typically involve large temperature gradients between adjacent half-turns, as the thermal conductivity of the superconducting cable is significantly larger than that of the surrounding electric insulation. Capturing these gradients accurately typically requires a high-quality, fully resolved mesh of the insulation layers~\cite[Sec.~5.3]{Monk2003}, leading to a large number of degrees of freedom (DoFs) and, consequently, high computational costs. For coupled FE simulations, two different meshes are often used for the thermal and magnetic models~\cite{Driesen2001}. By enforcing suitable interface conditions, the thermal thin shell approximation (TSA) provides an efficient alternative without the need for explicit meshing~\citep{Driesen2001, Schnaubelt2025, Schnaubelt2023a, Chan2010, Schnaubelt2024}. 

\begin{figure*}[ht]
    \centering
    \includegraphics[width=.49\linewidth]{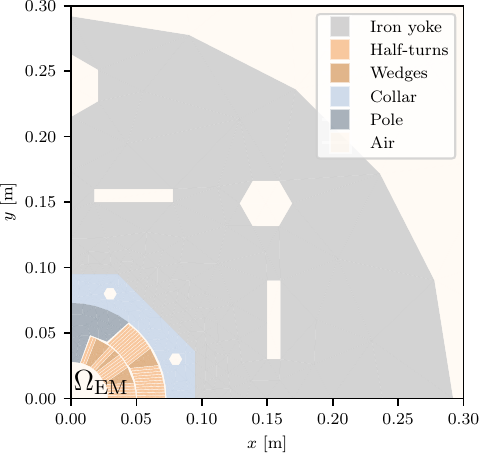}
    \includegraphics[width=.49\linewidth]{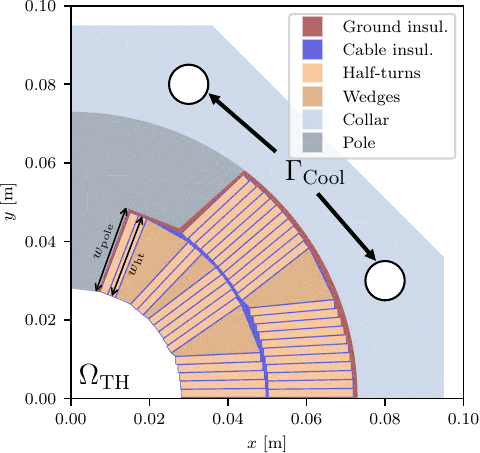}
    \caption{Illustration of the electromagnetic computational domain $\Omega_\mathrm{EM}$ and thermal computational domain $\Omega_\mathrm{TH}$ for a 12~T~dipole magnet design suggested in \cite{Foussat2026}. Only a quarter of the magnet is shown. The collar, pole, and ground insulation regions have been integrated into FiQuS. The materials are specified in Tab.~\ref{tab:materials}.}
\label{fig:dipole_regions}
\end{figure*}

While commercial tools like COMSOL Multiphysics can model thin thermal layers~\cite{Bortot2017}, these rely on proprietary, closed-source implementations. The detailed derivation of the TSA is instead presented and implemented in an open-source framework in~\cite{Schnaubelt2023a}. Despite these advances, current state-of-the-art models still typically neglect the thermal influence of structural components such as the pole and collar.

Cooling mechanisms are critical for stable magnet operation. Standard approaches for modelling immersive cooling methods include convective cooling boundary conditions via a heat transfer coefficient~\cite{Verweij2006}, e.g. in FiQuS~\cite[Sec.~2.2]{SchnaubeltPhD}, or by modelling a confined helium layer next to the insulation, as is done in ANSYS~\cite{Chorowski2006}. In the Large Hadron Collider (LHC) main dipoles, the non-impregnated Nb-Ti cables allow superfluid helium to permeate the coil and provide efficient cooling~\cite{Bottura2024}.  However, the amount of liquid helium required and its potential leakage can lead to a high cost.

Consequently, alternative cooling strategies are being explored. Conduction cooling has been investigated for magnetic resonance imaging scanners as a sustainable alternative for liquid helium immersion~\cite{Xu2025}. For future particle accelerators, conduction cooling through the collar region represents a promising alternative~\cite{Shimizu2022, Nezuja2026}. Although providing lower local cooling power than direct immersion, it has the potential to be easier to regulate, be more environmentally sustainable by reducing helium consumption, and lead to a simpler magnet design. In this work, the helium temperature is fixed at $1.9$\,K, as this is used in some high-field accelerator magnets, but this could potentially be increased to $4.5$\,K to further reduce operational costs~\cite{BorgesDeSousa2026}. High-field magnets often rely on impregnated \ce{Nb_3Sn} coils~\cite{Todesco2018}.  Since impregnation prevents helium from penetrating the coil pack for direct cooling~\cite{Bottura2024}, accurate modelling of conduction cooling is essential.
\begin{table}
\centering
\captionof{table}{Materials used for the regions in Fig.~\ref{fig:dipole_regions}.}
\label{tab:materials}
\includegraphics[width=\linewidth]{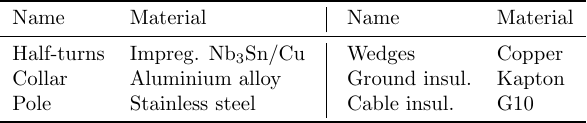}
\end{table}

The aim of this work is to extend the capabilities of FiQuS to incorporate the collar and pole regions, including the ability to have cooling channels within them. In particular, the novelty lies in connecting the coils, poles and collar via the thermal TSA. This includes determining how to deal with domains with non-conforming interfaces and, consequently, linking two separate TSAs with each other. While the TSA has proven efficient for quench simulations with high heat fluxes between magnet turns~\cite{Schnaubelt2025}, we aim to generalise its use to the collar and pole regions and evaluate its effectiveness in comparatively low heat fluxes between magnet turns. The novel method is applied to the dipole magnet design of a 12~T~dipole~\cite{Foussat2026}, shown in Fig.~\ref{fig:dipole_regions}, as a use case. To the best of our knowledge, this article presents the first FE model capable of simultaneously treating both transient magnetodynamic and thermal phenomena across the full magnet cross-section.

The paper is structured as follows. The magneto-thermal formulation and TSA are briefly summarised in Subsecs.~\ref{subsec:magnetothermal}~and~\ref{subsec:tsa} respectively. Particular attention is paid to the thermal coupling with the newly introduced regions. The effect of the meshing parameters is discussed in Subsec.~\ref{sec:Mesh_quality} for the fully resolved meshes and the TSA models. One TSA model is selected and compared to the fully resolved mesh models in terms of accuracy and computational speed in Subsec.~\ref{sec:TSA_performance}. The extended framework is applied to study several cooling configurations and collar materials in Subsec.~\ref{sec:Results_cooling}, to identify the maximal source current ramping rates without leading to a thermal runaway. Finally, we close this work in Sec.~\ref{sec:Conclusion} with some concluding remarks.

\section{Methodology}\label{sec:Methodology}
\subsection{Magneto-thermal formulation}\label{subsec:magnetothermal}
Magnetodynamic equations are considered in the electromagnetic computational domain $\Omega_\mathrm{EM}$ in an $\mathbf{A}$-formulation. The magnetic vector potential $\mathbf{A}$ is related to the magnetic flux density $\mathbf{B}~=~\nabla\times\mathbf{A}$ and the electric field $\mathbf{E}~=~-\tfrac{\partial{\mathbf{A}}}{\partial{t}}~+~\nabla~\phi$. The electric scalar potential $\phi$ can be eliminated by a gauge choice: $\mathbf{A}^*~=~\mathbf{A}~+~\int_{t_0}^t~\nabla \phi\,~\mathrm{dt}$, leading to the strong $\mathbf{A}^*$-formulation with the governing differential equation~\cite{DeGersem2020}~\cite[Sec.~5.18]{jackson1998}:
\begin{align}
\nu \nabla \times \mathbf{B}&= \mathbf{J}_s+\sigma \mathbf{E} \;, \nonumber \\
   \Leftrightarrow\;\; \nu \nabla \times (\nabla \times \mathbf{A}) + \sigma \tfrac{\partial \mathbf{A}}{\partial t} &= \mathbf{J}_s \;, \label{eq:strong_EM}
\end{align}
where $\nu=\nu(|\mathbf{B}|)$ denotes the magnetic reluctivity, $\sigma=\sigma(|\mathbf{B}|,~T)$ the electrical conductivity depending on the temperature $T$, and $\mathbf{J}_s$ the source current density. Since there are no applied currents outside the half-turns region $\Omega_\mathrm{ht}$, we have $\mathbf{J}_s=0$ in $\Omega_{\mathrm{EM}}\setminus\Omega_\mathrm{ht}$. The eddy current contribution is represented by the $\sigma\tfrac{\partial\mathbf{A}}{\partial t}$ term. 

For the half-turns regions, the multi-filamentary wires exhibit complicated cross-sections. This leads to complex inter-filament and inter-strand coupling currents as well as eddy currents in the stabilizer matrix~\cite{CAMPBELL1982}. A homogenisation approach can be introduced to account for such interactions between field and to reproduce magnetization, AC losses and voltages \cite{Dular2025, Glock2025, Dular2025b}. However, in this work, these effects are neglected. We apply Eq.~\eqref{eq:strong_EM} in $\Omega_\mathrm{ht}$, set $\nu=\nu_0$ and neglect the eddy current contribution in $\Omega_\mathrm{ht}$ for simplicity.

The thermal problem defined on the thermal computational domain $\Omega_\mathrm{TH}$ with boundary  $\partial \Omega_\mathrm{TH}$ relates the temperature $T$ to a heat source in the strong formulation as follows~\cite{Schnaubelt2025}:
\begin{align}
  P+ Q_\mathrm{ht} &= -\nabla\!\cdot\!{(\kappa \nabla T)}+C_V\partial_tT
  & \mathrm{in}\; \Omega_\mathrm{TH}, \label{eq:strong_TH} \\
  \vec{n}\!\cdot\!(\kappa\nabla T) &= h(T_\mathrm{He}\!-\!T) 
  & \mathrm{on}\; \Gamma_{\mathrm{Cool}}\!\subset\!\partial\Omega_{\mathrm{TH}}, \label{eq:Robin_bdry} \\
  \vec{n}\!\cdot\!(\kappa\nabla T) &= 0 
  & \mathrm{on}\; \partial \Omega_\mathrm{TH}\!\setminus\! \Gamma_\mathrm{Cool}. \label{eq:Neu_bdry}
\end{align}
where $\kappa=\kappa(|\mathbf{B}|,~T)$ denotes the thermal conductivity, $C_V=C_V(T)$ the volumetric heat capacity, $\vec{n}$ the normal to the boundary, and $Q_\mathrm{ht}$ the additional Joule losses in the half-turns, which are triggered when the local current density exceeds the current sharing threshold. This corresponds to Eq.~(3) in~\cite{Schnaubelt2025}. The power density $P=\sigma|\mathbf{E}|^2$ is a modified, ramp rate dependent heat source, introduced to include a non-zero heat source below current sharing regime. Both the heat transfer coefficient  $h=h(T)$ and the temperature of the liquid helium $T_\mathrm{He}=1.9$\,K are used in Eq.~\eqref{eq:Robin_bdry} to define a Robin-type boundary condition for the cooling holes. The last equation imposes adiabatic conditions, i.e. no heat can escape through $\partial\Omega_\mathrm{TH}\!\setminus\!\Gamma_\mathrm{Cool}$. The half-turns region is modelled as a single homogeneous domain, where the different material properties are accounted for via volumetric fractions $f$, e.g. for $\sigma|_{\Omega_\mathrm{ht}}=f_\mathrm{\ce{Cu}}\sigma_{\ce{Cu}}$, as proposed~in~\cite[Eqs.~(1), (2) and~(4)]{Schnaubelt2025}.

Using a Ritz–Galerkin approach, the strong formulations Eqs.~\eqref{eq:strong_EM}-\eqref{eq:Neu_bdry} are converted into weak formulations by introducing test functions and integrating over the spatial domain, as in~\cite{Schnaubelt2023a, SchnaubeltPhD, DeGersem2020}.

To reduce computational cost, typically not all magnet parts are modelled, as is also visible in Fig.~\ref{fig:dipole_regions}. Due to the high temperature gradients resulting from low thermal conductivity, the insulation requires a fine thermal mesh. In contrast, the electromagnetic simulation treats these regions as an air domain, allowing for a coarser discretization. This leads typically to the use of different meshes for each simulation type. Information is transferred through a projection from one system to another~\cite{Parent2008}, which is required since the material properties are functions of $\mathbf{B}$~and~$T$.

\begin{figure}
    \centering
    \includegraphics[width=\linewidth]{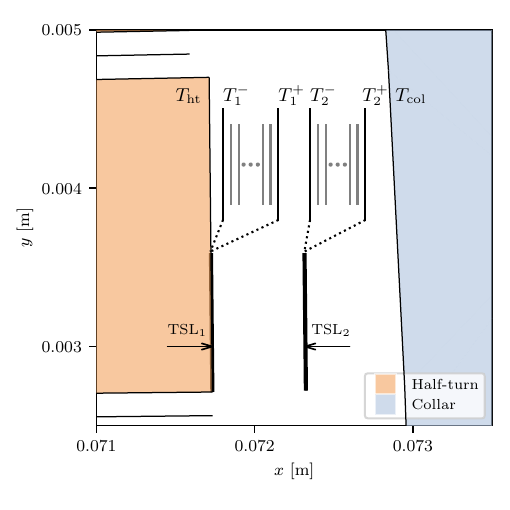}
    \caption{Visualisation of the collar TSA and its internal discretisation. The lengths are not to scale for readability. The first and second TSLs replace the cable and ground insulation respectively, as depicted in Fig.~\ref{fig:tsa}.}
    \label{fig:internal_TSA}
\end{figure}

\subsection{Thin shell approximation}\label{subsec:tsa}
\subsubsection{General idea and closest neighbour mapping}
\begin{figure*}[ht]
    \centering
    \begin{minipage}[b]{0.48\linewidth}
        \centering
        \includegraphics[width=\linewidth]{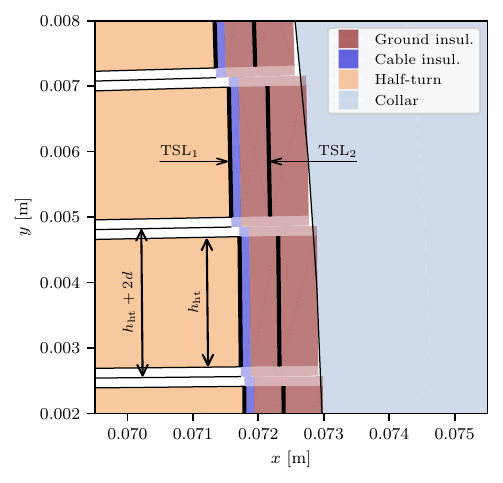}
        \caption{Visualisation of the modelled material area in the TSA for the collar. TSL$_1$ and TSL$_2$ are indicated by thicker lines. The darker region represents the model without length correction.}
        \label{fig:tsa}
    \end{minipage}
    \hfill
    \begin{minipage}[b]{0.48\linewidth}
    \centering
    \includegraphics[width=\linewidth]{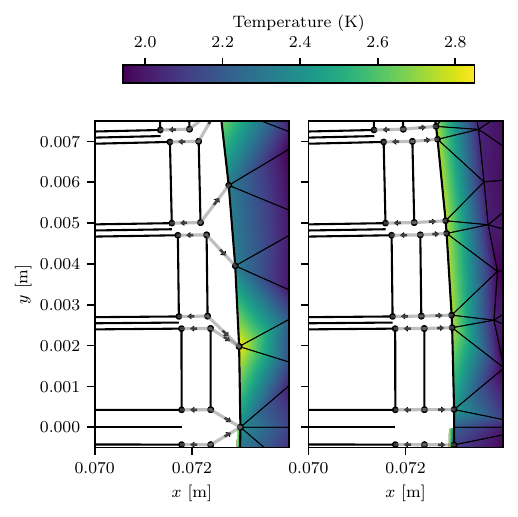}
    \caption{Illustration of a free closest neighbour map (left) and an enforced map (right). For clarity, only the collar temperature is shown at a specific point in time, for a quenching simulation and a stainless steel collar. Note the difference in local temperatures.}
    \label{fig:coord_map_and_temp}
    \end{minipage}
\end{figure*}

In the two-dimensional model, thin shell lines (TSLs) are employed to replace the insulation layers between half-turns, wedges, layers, poles and the collar. Instead of meshing these thin regions as a surface, the TSLs impose an interface condition that introduces a thermal discontinuity across the insulation. This leads to geometrically disconnected domains, and consequently, more meshing freedom as the meshing would otherwise be constrained by the insulation surface mesh. 

The TSL has an internal discretisation, which is obtained by a tensor-product based FE method~\cite{Schnaubelt2023a}, allowing the TSL to represent a temperature discontinuity as shown in Fig.~\ref{fig:internal_TSA}. Heat can be conducted in 
the tangential direction within one TSL, but not between two disconnected TSLs. Furthermore, the insulation thickness along a TSL is considered constant, such that the line can be treated as a 1D representation of a rectangular surface, depicted in Fig.~\ref{fig:tsa}. This approximation has previously been applied for dipole magnets in~\cite{Schnaubelt2025}, with the exception of the interface involving the collar. However, in the latter, only the coil pack itself was considered. This work adds collar and pole regions to allow for more comprehensive thermal simulations. 

For our model problem, there are two different insulation materials, which are modelled by two TSLs. The outer sides of the half-turns incorporate the cable insulation with a first TSL located on the half-turn boundary, illustrated by TSL$_1$ in Fig.~\ref{fig:internal_TSA}. The link constraint matches the half-turn temperature ($ T_\mathrm{ht}=T^\mathrm{-}_1$) on one side. Next, the ground insulation material between the half-turn and the collar is modelled by a second TSL, indicated by TSL$_2$, which is constrained by TSL$_1$ ($T^\mathrm{+}_{1}=T^-_2$) and the temperature in the collar ($T^\mathrm{+}_2 = T_\mathrm{col}$). Similarly to what is proposed in \cite{Schnaubelt2025}, the temperature of the collar or pole mesh node closest to the TSL is used, illustrated by TSL$_2$ in Fig.~\ref{fig:coord_map_and_temp}. 

For quench simulations or when using a low thermal conducting collar material, this closest neighbouring mesh node mapping might cause numerical artefacts in the simulation. This is illustrated by Fig.~\ref{fig:coord_map_and_temp}. On the left figure, two links are associated with the same mesh node, and since the total heat flux is conserved, there is an increase in local temperature compared to the figure on the right. This can lead to large discrepancies in maximum collar temperature. The problem can be avoided by applying a mortaring method to link both sides of a non-conformal mesh, enforcing equal heat flux by adding Lagrange multipliers~\cite{Hahn2024}. This comes at the cost of additional unknowns and consequently a higher computational complexity. 

Instead, a small mesh manipulation is implemented which enforces the mesh node position on the collar side to guarantee a one-to-one mapping. As illustrated by Fig.~\ref{fig:coord_map_and_temp}, this removes the artificial local differences in temperature, but compromises the collar mesh freedom. 

\subsubsection{Appropriate length scaling}
As mentioned in the previous section, the TSA assumes that the tangential heat flux between different TSLs in the insulation layer is negligible. For simulations with fast quenching half-turns, this assumption leads to accurately computed maximum temperatures~\cite{Schnaubelt2025}. However, one must be cautious when generalising this. First of all, the average insulation thickness between the collar and half-turns or wedges (ground insulation in Fig.~\ref{fig:dipole_regions}) is considerably larger than in-between half-turns and wedges (cable insulation in Fig.~\ref{fig:dipole_regions}). As a result, more internal discretization levels in the TSA are needed.  Moreover, the half-turn width ($w_\mathrm{ht}$ in Fig.~\ref{fig:dipole_regions}) is typically much smaller than its height ($h_\mathrm{ht}$ in Fig.~\ref{fig:tsa}).

Taking into account the cable insulation with a thickness~$d$, we make a distinction between lines along the radial and azimuthal direction. The former, appearing between one half-turn and the collar or pole, are assumed to have a contact length about $15\%$ larger ($[h_\mathrm{ht}+2d]/h_\mathrm{ht}=1.1508$, see Fig.~\ref{fig:tsa}) for a 12~T~dipole magnet. Similarly, the azimuthal lines, only appearing at the poles, represent a contact length which is $7\%$ larger ($w_\mathrm{pole}~/~w_\mathrm{ht}~=~1.0701$, see Fig.~\ref{fig:dipole_regions}). 

We use a scaling factor in order to diminish the error introduced by these geometric differences for the thin shell length, without compromising the computational speed by adding additional calculations or restraining the mesh. A similar scaling of material properties has previously been applied in~\cite{Schnaubelt2024} for pancake coils. In our case, we model the TSL as an extension of the half-turns (thus maintaining the bare half-turn length) and scale the material parameters with a factor, effectively modelling a longer thin shell, as depicted in Fig.~\ref{fig:tsa}. The main benefit of keeping the thin shell length the same as the half-turns on a geometrical level is for the mapping and for the mesh enforcement explained earlier. Concerning the TSLs connecting the wedges to the collar and pole regions, no correction is applied, assuming the insulation thickness is very thin compared to the wedge or half-turn length. This scaling law can only partially cover the effect of the tangential heat flow and it does not account for the mid-pole additional insulation. Moreover, since a linear scaling is applied to non-linear material properties, it cannot match the reference model perfectly. However, the approach leads to reasonable accuracy as will be shown in Sec.~\ref{sec:Results}.

\subsection{Reference solution}\label{subsec:full_mesh}
In order to verify the TSA, a fully resolved reference mesh is used. Since the temperature gradients are expected to be relatively small compared to quench simulations, a coarse mesh could potentially be feasible. Therefore, different mesh sizes are considered in Sec.~\ref{sec:Mesh_quality}. However, automating such a surface meshed workflow is considerably more involved, as the insulated regions are not explicitly defined and require additional meshing logic.

\section{Results and discussion}\label{sec:Results}
We focus on a current ramp-up simulation including Joule losses from induced eddy currents that heat up the coils, wedges, collar and pole regions. We used a relatively fast ramp rate to verify the TSA for temperatures closer to the current sharing temperature of the winding. This is to confirm that the added functionality is suitable for the ramp rate sensitivity study of the magnets when they are ramped at ramp rates which cause quench below the nominal currents. 

While the method is also applicable to quench simulations, such cases are less informative for benchmarking the new thermal connections, since the dominant heat transfer between half-turns and wedges has already been extensively validated, e.g. in~\cite{Schnaubelt2025}. The ramp-up scenario therefore provides a more discriminating test case for evaluating the added value of the novel TSA including the collar and pole regions.

As mentioned in Sec.~\ref{sec:Methodology}, the simulations utilize distinct meshes for the thermal and electromagnetic domains. To isolate the effects of the TSA in the following analysis, the electromagnetic mesh is kept identical across all models. 

\begin{figure}
    \centering
        \centering
        \includegraphics[width=\linewidth]{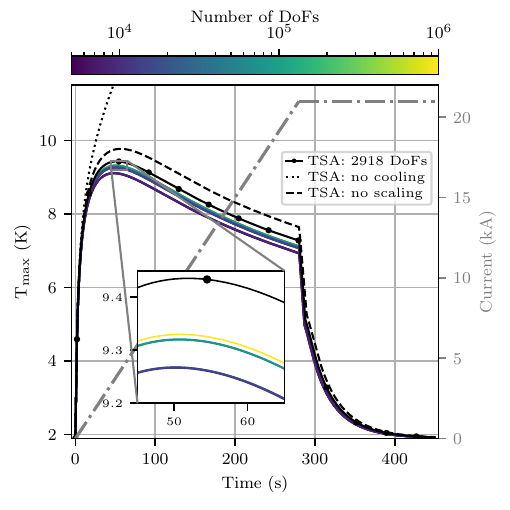}
        \caption{Maximum temperature during a current ramp (gray) for fully resolved mesh with different mesh sizes (colour bar) and a TSA model (black). TSA models without scaling and no cooling are also plotted for comparison. Mesh size parameters are specified in Tab.~\ref{tab:reference_mesh_table} and Fig.~\ref{fig:relative_error} for the TSA.}
    \label{fig:reference_mesh_study_TMAX}
\end{figure}

\subsection{Mesh quality: parameter study}\label{sec:Mesh_quality}
\subsubsection{Fully resolved mesh models}\label{sec:Mesh_quality_REF}

The mesh quality is analysed by varying the mesh sizes, i.e. the characteristic size of the mesh triangles, and comparing the simulation results to those obtained with the most refined mesh. We consider variation of the insulation mesh and collar mesh sizes. The mesh size of a region will constrain the meshing of the neighbouring regions as conformal meshes are required. The coil mesh could be considered, but since it will be mostly isothermal, it should not require a fine mesh and further constrain the mesh of other regions. The reported relative errors are with respect to the most refined fully resolved mesh, shown in yellow in Fig.~\ref{fig:reference_mesh_study_TMAX}.

First, looking at the maximum temperature $T_\mathrm{max}$ on $\Omega_\mathrm{TH}$ as a function of time $t$ in Fig.~\ref{fig:reference_mesh_study_TMAX}, we can see a combination of the half-turns heating up due to the transient effects, and the cooling which is more pronounced at higher temperatures due to the Robin condition. For comparison, one simulation is shown (dotted) without cooling, in which a thermal runaway occurs. In contrast, for the cooled simulations, these competing effects result in a maximum temperature of around $9$\,K for this simulation at around $60$\,s. Then, the system slowly cools down until the full current is reached (at $280$\,s) and the transient effects fade away. The magnet then cools back down to $T_\mathrm{He}$. While doing so, the maximum temperature can move to a different half-turn, which may be cooling down more slowly leading to `kinks' in the curve, e.g. around $290$\,s.  

\begin{figure}
        \centering
        \includegraphics[width=\linewidth]{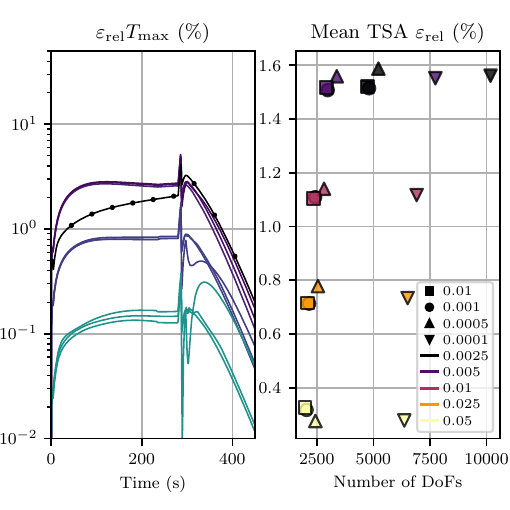}
        \caption{(left) The relative error of the maximum temperature compared to the reference model for various mesh sizes (colour bar of Fig.~\ref{fig:reference_mesh_study_TMAX}) and the TSA model with 2918 DoFs (black). (right) Time averaged relative error of TSAs using different collar mesh sizes (colour, in meters) and TSA discretization sizes of the insulation (marker, in meters). The pole mesh size is half the collar mesh size.}
        \label{fig:relative_error}
\end{figure}

The relative error of $T_\mathrm{max}$, $\varepsilon_\mathrm{rel}T_\mathrm{max}$, with respect to a reference mesh, decreases with increasing number of DoFs, see Fig.~\ref{fig:relative_error} (left). This is mainly 
related to the insulation mesh size, which is specified in Tab.~\ref{tab:reference_mesh_table}. A peak in the relative error is observed at the end of the current ramp, after which artefacts in the relative error appear since the location of the maximum temperature changes over time. These artefacts are a consequence of the chosen error metric and can be safely disregarded. 

We further noticed that the collar and pole mesh size do not influence the maximum temperature. The only noticeable difference is the local maximum temperature in the pole and collar region itself. This is not of great importance since aluminium alloy is very thermally conducting, and the different mesh sizes all result in low maximum temperatures with respect to the half-turn temperatures. Hence, this yields a similar cooling effect on the half-turns. This can be seen from the models with a coarser collar mesh which are still able to capture the maximum temperature accurately. However, when modelling a stainless steel collar, the collar exhibits larger temperature gradients and the collar mesh is more important to correctly describe the cooling.

\subsubsection{TSA mesh}\label{subsec:tsa_mesh}

\begin{table}
\centering
\captionof{table}{Computation times for the assembly of the system matrices and the solution of fully resolved simulations using different thermal mesh resolutions, while the electromagnetic mesh is kept identical across all models. For comparison, the TSA model is added in gray and models mentioned in the text are marked bold and shown in Fig.~\ref{fig:meshes}. The number of DoFs correlates with computation time and accuracy, as shown in Figs.~\ref{fig:reference_mesh_study_TMAX} and~\ref{fig:relative_error}. Rows are loosely sorted by performance. CPU: Intel Xeon E5-2667 v4 @ 3.20GHz with 16 cores.}
\label{tab:reference_mesh_table}
\includegraphics[width=\linewidth]{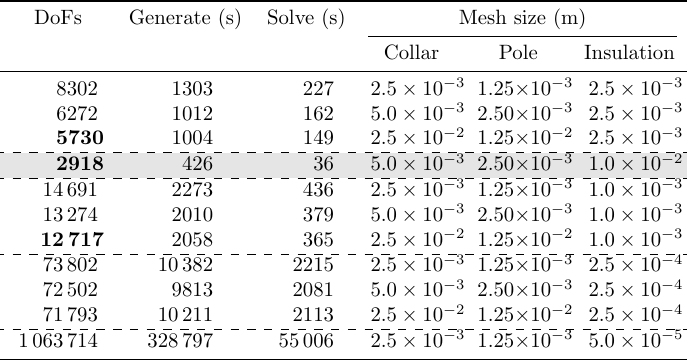}
\end{table}

Next, we look at TSA models and the effect of the meshing parameters on the TSA solution. Unlike the fully resolved meshes, the TSA models employ quadrilateral elements to mesh the half-turns and wedges, as they are not constrained by a surrounding mesh and are approximately isothermal. Fig.~\ref{fig:relative_error} (right) illustrates the mean error of a simulation for different collar and pole mesh sizes (assuming pole mesh size is half the collar mesh size) and different insulation mesh sizes, which determines the number of internal TSLs. The TSA is compared against the most refined fully meshed model. 

By varying the meshing parameters for the TSA, we observed that the solution does not further converge toward the fully resolved reference mesh. This behaviour can partly be attributed to the modelling assumptions, as the two models are not necessarily expected to yield identical results. The discrepancy is likely related to the assumption of negligible tangential heat flux in the insulation layer. While a more sophisticated scaling factor might improve the accuracy, finding a general solution may be cumbersome.

\begin{figure*}[ht]
    \centering
    \includegraphics[width=.325\linewidth]{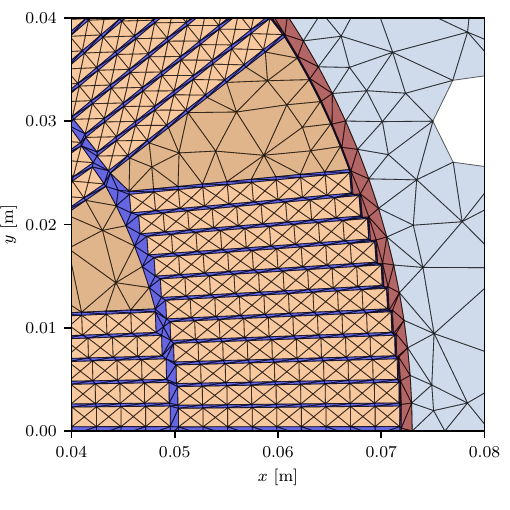}
    \includegraphics[width=.325\linewidth]{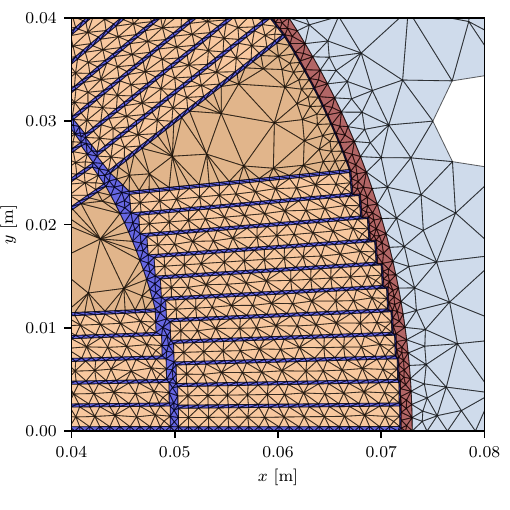}
    \includegraphics[width=.325\linewidth]{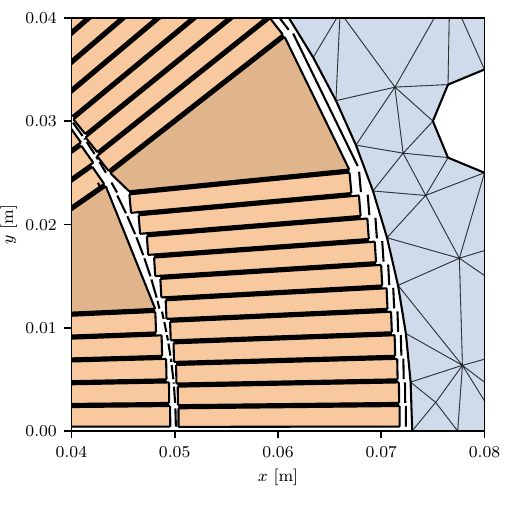}
    \caption{Fully resolved thermal meshes with 5730 (left) and 12\,717 (center) DoFs used for comparison with the TSA (right) with 2918 DoFs, highlighted in bold in Tab.~\ref{tab:reference_mesh_table}. Colour legend is shown in Fig.~\ref{fig:tsa}.}\label{fig:meshes}
\end{figure*}

Based on Fig.~\ref{fig:relative_error}, it can be seen that some TSA models yield errors much below $1\%$, however, the more refined TSA models yield about $1.6\%$ mean error. It can also be noted that the changing discretisation of the insulation does not alter the average error, meaning that few internal discretisations are sufficient to resolve the temperature gradient across the insulation layer. We conclude that the mean error depends mainly on the collar and pole mesh sizes. For this reason, we select a single TSA model for the remainder of the discussion. This model has sufficient mesh resolution while using the smallest number of DoFs, approximately~3\,k.

\subsection{TSA performance}\label{sec:TSA_performance}
From Fig.~\ref{fig:reference_mesh_study_TMAX} we note that TSA with scaling is closer to the reference model compared to the TSA without scaling, thus highlighting the necessity of introducing the scaling factor. Compared to the fully resolved reference mesh in yellow, Fig.~\ref{fig:reference_mesh_study_TMAX} also shows that the TSA overestimates the temperature, while the simulations with a fully resolved mesh are persistently underestimating the reference model.

A quench occurs if the source current in the half-turns surpasses the critical current determined by the half-turn temperature and the local magnetic field. This leads to a rapid increase in temperature. For this reason, employing the TSA for maximum ramp rate estimation is considered a more conservative approach, as it predicts a quench at lower ramp rates. In contrast, the use of a coarse mesh may fail to predict a quench.

We conclude that the TSA reaches accuracies below $2\,\%$ during and below $4\,\%$ after the ramp. This is slightly worse than the fully resolved mesh with $13$\,k DoFs, with errors below $0.8\,\%$ and up to $5\,\%$, but better than the $6$\,k DoFs model. Regarding the computational cost in Tab.~\ref{tab:reference_mesh_table}, we can see a reduction in total generation and solving time for the TSA model. It should be noted that the total computation time for every model includes the EM system, which requires about 72\,s for generation and 15\,s for solving. As wall generation time scales approximately linearly with the number of DoFs, while the wall solving time scales superlinearly, the TSA leads to significantly faster simulations. The TSA yields a total speed-up between $250\%$ and $517\%$, compared to the models marked in bold in Tab.~\ref{tab:reference_mesh_table} and meshes shown in Fig.~\ref{fig:meshes}. Although the observed speed-up in this work is smaller than for quench simulations reported in~\cite{Schnaubelt2025}, this difference is expected. In~\cite{Schnaubelt2025}, simulations require significantly finer meshes in the insulation layer to resolve steep temperature gradients and large heat fluxes. Nevertheless, the TSA still achieves a noticeable speed-up for the ramping simulations considered here.

In 2D simulations, generation times can become larger than solution times. In this context, a dedicated assembly procedure has been implemented in GetDP 4.0~\cite{GetDP} to efficiently handle the large number of terms in the weak formulations associated with each TSL. In particular, the current model description may lead to some TSA meshes having more weak formulation terms than DoFs. Special care is taken to perform basic elementary operations, such as element-support mapping, just once during the very first assembly. This approach avoids the significant overhead otherwise observed during subsequent assemblies for these application-specific edge cases, as it reduced the generation time reported in Tab.~\ref{tab:reference_mesh_table} by a factor of 9 for the 3\,k DoFs model.

\subsection{Example: cooling hole position study}\label{sec:Results_cooling}
The developed model is now employed to assess cooling strategies in the collar. Only the TSA models are used since fully meshed models are too costly. Cooling is implemented via the Robin boundary condition~\eqref{eq:Robin_bdry} at the boundaries of the cooling holes, denoted as $\Gamma_\mathrm{Cool}$ in Fig.~\ref{fig:dipole_regions}.

\begin{figure}
    \centering
    \includegraphics[width=\linewidth]{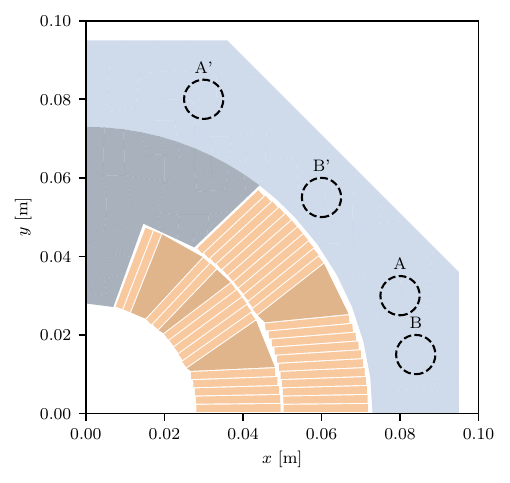}
    \caption{Cooling hole configurations $A$ and $B$. Each configuration contains two distinct cooling holes, denoted $X$ and $X'$. Only a quarter of the magnet is shown; the remaining three quarters follow by symmetry (compare Fig.~\ref{fig:cooling_configs_nbrs}).}
    \label{fig:cooling_configs_holes}
\end{figure}

\begin{figure}
    \centering
    \includegraphics[width=\linewidth]{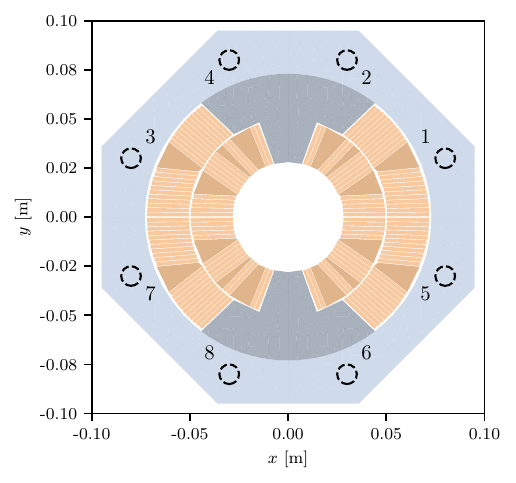}
    \caption{The numbering for the cooling holes in configuration $A$. In configuration $A_o$ ($A_e$), only the odd (even) numbered cooling holes are active, while in $A_{1,7}$, cooling is provided through holes 1 and 7.}
    \label{fig:cooling_configs_nbrs}
\end{figure}

Several configurations, varying in the number and positions of the cooling holes, are considered. Configuration $A$ compromises eight cooling holes, as shown in Fig.~\ref{fig:cooling_configs_holes}, while configuration $B$ also consist of eight cooling holes, but positioned closer to the half-turns. Fig.~\ref{fig:cooling_configs_nbrs} shows the hole numbering in configuration $A$, with $A_o$ and $A_e$ indicating that only odd- and even-numbered holes are active respectively. The two-hole configuration $A_{1,7}$, equivalent to $A_{3,5}$ by symmetry, is selected to test two-hole cooling.

\begin{table}
\caption{Lower and upper bounds on the maximal ramp rate considering different cooling configurations (Figs.~\ref{fig:cooling_configs_holes} and~\ref{fig:cooling_configs_nbrs}) and collar materials. The increase in collar temperature with respect to the initial temperature of $1.9$\,K is reported for the lower bound (i.e. non-quenching) simulations.}\label{tab:cooling_rates}
\centering
\includegraphics[width=\linewidth]{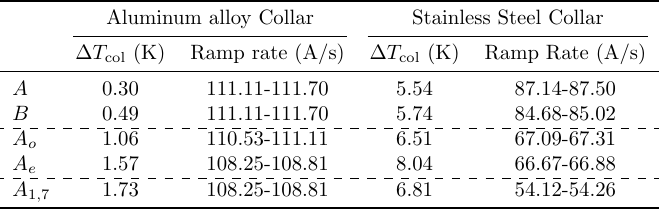}
\end{table}

In order to find the fastest ramp rate without thermal runaway, an upper and lower bounds are estimated by varying the rates until a quenching and non-quenching simulation is found. For the non-quenching simulation, the increase in collar temperature is reported. The ramp time is discretized in fixed increments of $1$\,s, the corresponding ramp rates are reported as ranges.

Concerning the aluminium alloy collar, the ramp rates reported in Tab.~\ref{tab:cooling_rates} are roughly identical for all cooling positions; with configurations $A_e$ and $A_{1,7}$ requiring the slowest ramp rate. Looking at the increase in collar temperature, which is much smaller than for stainless steel, configuration $A$ is able to keep the collar the coldest. Note that configuration $A_o$ only uses 4 cooling holes, but performs only slightly worse than configuration $A$. To conclude, for the aluminium alloy collar the cooling positions are of little importance, and the use of only two cooling holes ($A_{1,7}$) may be considered. 

Acceptable ramp rates for the stainless steel collar are much lower, which is expected due to the reduced thermal conductivity compared to aluminium alloy. As the heat distributes much slower, the positioning of the cooling holes is more crucial. The heat capacity of steel is also lower, which means that it requires less heat to increase its temperature. This is visible from the increase in collar temperature, which can be explained by these effects. Configuration $A$ shows a small advantage over configuration $B$, allowing for a slightly faster ramp. Configurations $A_o$ and $A_e$ are comparable, but considerably slower than cooling with 8 cooling holes.

\begin{figure*}
    \centering
    \includegraphics[width=0.33\linewidth]{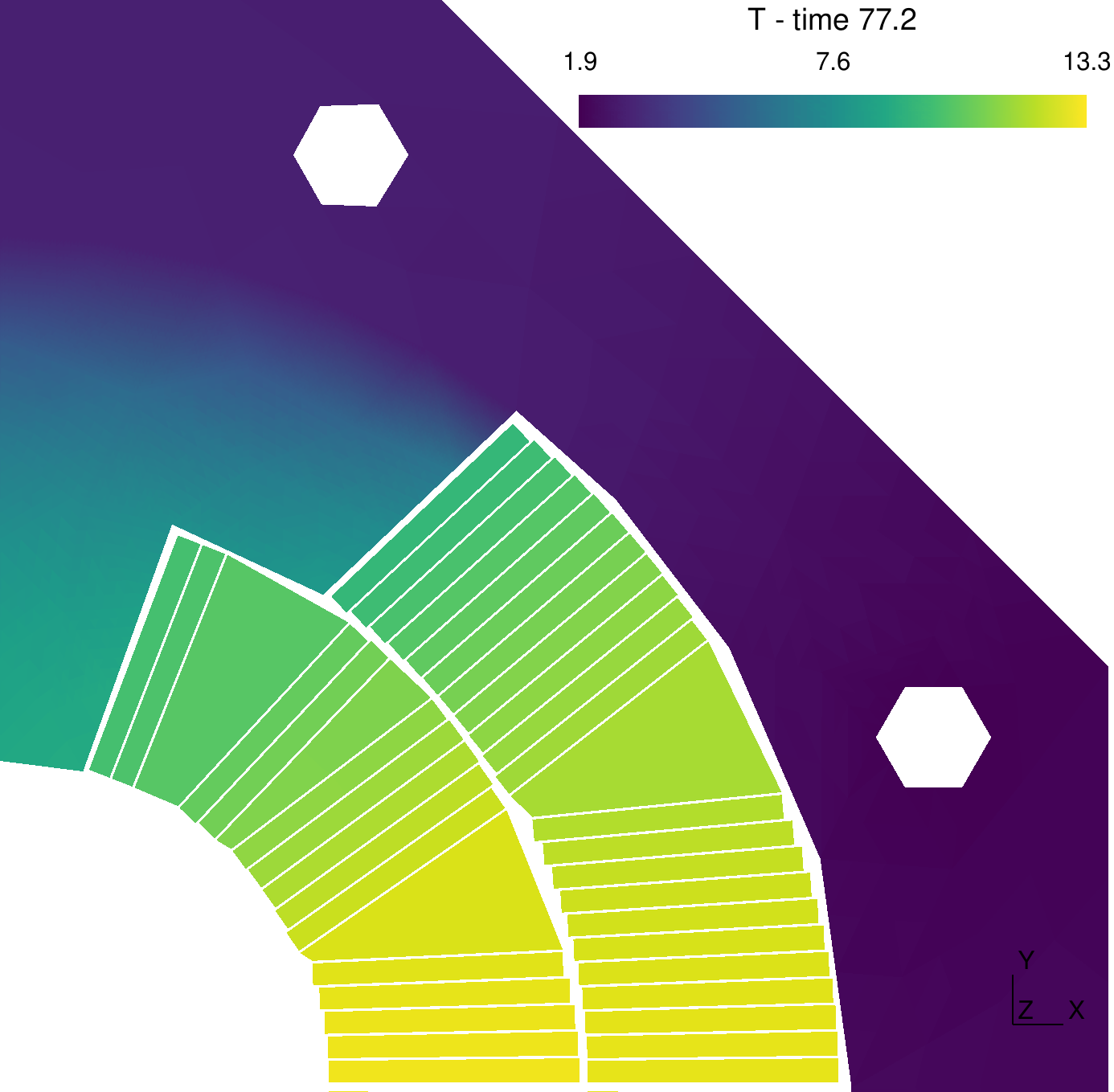}
    \hspace{5em}
   \includegraphics[width=0.33\linewidth]{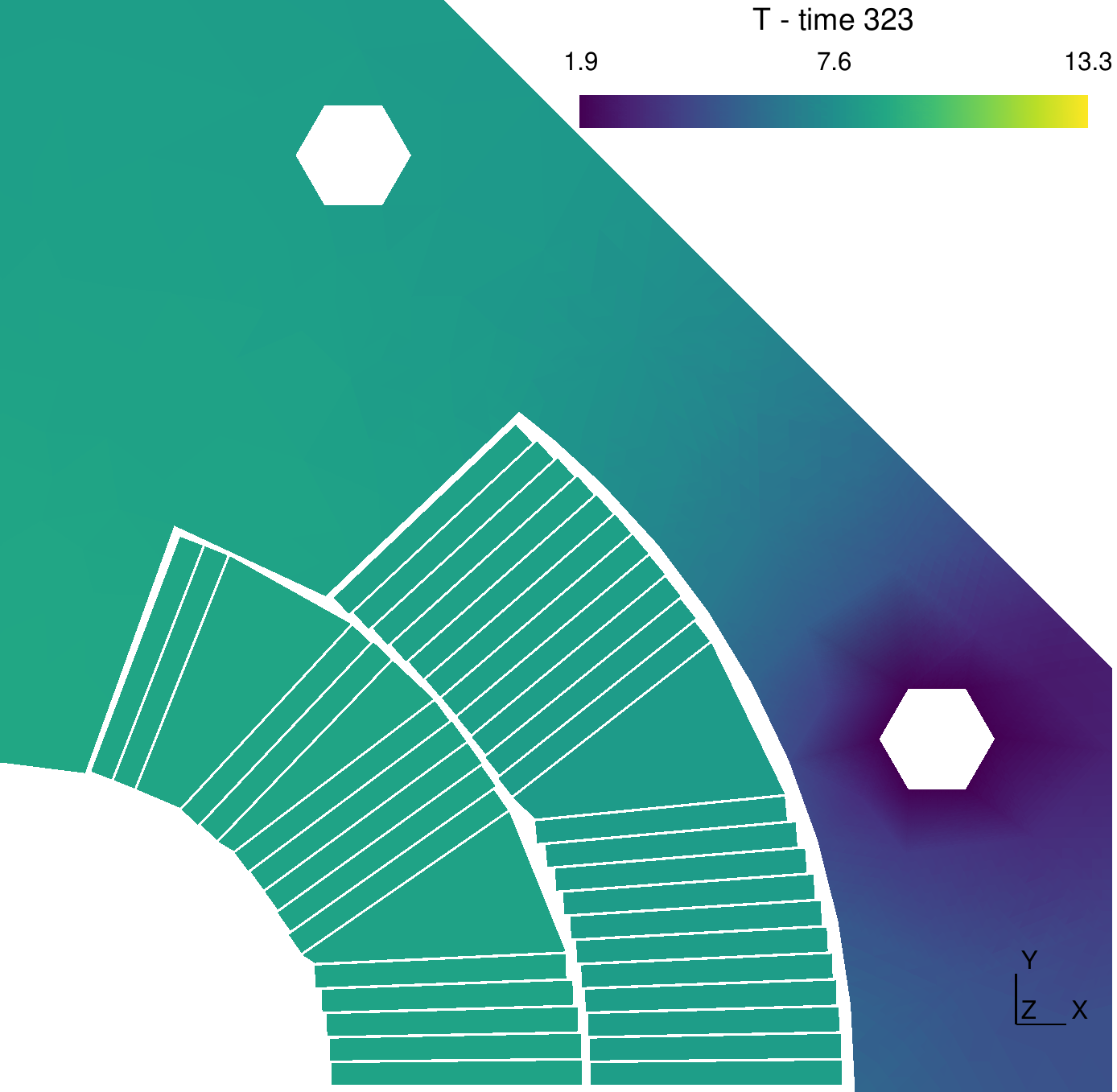}
    \caption{Temperature distribution (in K) for the $A_o$ cooling configuration with an aluminium alloy collar and ramp rate of $110.52$\,A/s (left), and a stainless steel collar and ramp rate of $67.09$\,A/s (right), each at the time (in s) when $\Delta T_\mathrm{col}$ in Tab.~\ref{tab:cooling_rates} is reached.}
    \label{fig:temperature_distribution}
\end{figure*}

The temperature distribution for the $A_o$ configuration is shown in Fig.~\ref{fig:temperature_distribution} for both the aluminium alloy and stainless steel collar, at the time when $\Delta T_\mathrm{col}$ is reached. 

To conclude, the inclusion of the novel collar and pole regions within the FiQuS framework enables the systematic study of conduction cooling configurations of future accelerator magnets. However, a complete and physically accurate modelling of AC loss contributions in the half-turns is not yet included and will be addressed in future work.

~

\newpage
\section{Conclusion}\label{sec:Conclusion}
Reliable and accurate transient simulations are important for improving the protection and operational safety of superconducting accelerator magnets. In this contribution, we presented the FiQuS tool which has been extended to incorporate the collar and pole regions, allowing for coupled magneto-thermal simulations on a more complete magnet cross-section. To the best of our knowledge, this represents the first FE model capable of simultaneously treating transient magnetodynamic and thermal effects with the pole and collar regions included.

To achieve this, we generalised the TSA to the collar and pole interfaces, thereby avoiding the need for explicit meshing of the insulation layers. This increases meshing flexibility, while maintaining an acceptable level of accuracy. In this work, a necessary length correction was introduced to account for the tangential heat flux between neighbouring thin shell layers. This was then applied to the simulation of a 12~T~dipole magnet as a use case. Tangential and radial contributions are treated separately, such that the correction can be generalised to other magnets. Additionally, an automated workflow for the TSA is implemented, which would be substantially more difficult to construct in the case of fully resolved mesh models. 

The TSA models were shown to reproduce the temperature evolution of the fully resolved mesh reference model of a 12~T~dipole magnet within a relative error of approximately 2–4\,\%. The computational time required is nevertheless reduced by up to a factor 5. Although the TSA does not converge towards the reference for finer mesh discretisations, it provides a robust and efficient alternative for estimating maximum allowable ramp rates. Furthermore, the method remains applicable under both quenching and steady-state conditions. 

As an illustration, the TSA model was used to evaluate cooling strategies for the collar region. It was shown that aluminium alloy collars were largely insensitive to the cooling hole configuration, unlike stainless steel collars. This shows that aluminium alloy collars can improve thermal performance, allowing for similar current ramp rates with fewer cooling holes.

Future work will focus on including the AC loss of superconducting coils using a homogenisation approach. Additionally, the TSA accuracy could be improved by refining the TSA length scaling.

\section*{Code Availability Statement}
All scripts and source code required to reproduce the simulations presented in this work are publicly available at \url{https://gitlab.cern.ch/steam/analyses/fiqus-collar-and-pole-tsa-verification-study}\,.
\section*{CRediT authorship contribution statement}
\textbf{E.V.} Conceptualisation, Methodology, Software, Validation, Visualization, Writing—original draft, Writing—review and editing \textbf{E.S.} Conceptualization, Methodology, Software, Writing – review and editing, Supervision \textbf{L.D.} Software, Writing – review and editing \textbf{C.G.} Software, Writing – review and editing \textbf{A.V.} Writing – review and editing  \textbf{M.W.}  Conceptualization, Writing – review and editing, Supervision.
\section*{Declaration of competing interest}
The authors declare that they have no known competing financial interests or personal
relationships that could have appeared to influence the work reported in this paper.
\section*{Acknowledgements}
This work was partially supported by the CERN High-Field Magnet (HFM) programme. The work of L.D. was supported by the F.R.S.-FNRS.

\bibliographystyle{elsarticle-num}
\bibliography{bib}

\begin{thebibliography}{10}
\expandafter\ifx\csname url\endcsname\relax
  \def\url#1{\texttt{#1}}\fi
\expandafter\ifx\csname urlprefix\endcsname\relax\def\urlprefix{URL }\fi
\expandafter\ifx\csname href\endcsname\relax
  \def\href#1#2{#2} \def\path#1{#1}\fi

\bibitem{Bortot2018}
L.~Bortot, B.~Auchmann, I.~C. Garcia, A.~M.~F. Navarro, M.~Maciejewski, M.~Mentink, M.~Prioli, E.~Ravaioli, S.~Schps, A.~P. Verweij, {STEAM: A Hierarchical Cosimulation Framework for Superconducting Accelerator Magnet Circuits}, IEEE Transactions on Applied Superconductivity 28~(3) (2018) 1--6.
\newblock \href {https://doi.org/10.1109/TASC.2017.2787665} {\path{doi:10.1109/TASC.2017.2787665}}.

\bibitem{steam_website}
{CERN STEAM Team}, {STEAM: Simulation of Transient Effects in Accelerator Magnets}, \url{https://cern.ch/steam} (2023).

\bibitem{Vitrano2023}
A.~Vitrano, M.~Wozniak, E.~Schnaubelt, T.~Mulder, E.~Ravaioli, A.~Verweij, {An Open-Source Finite Element Quench Simulation Tool for Superconducting Magnets}, IEEE Transactions on Applied Superconductivity 33~(5) (2023) 1--6.
\newblock \href {https://doi.org/10.1109/TASC.2023.3259332} {\path{doi:10.1109/TASC.2023.3259332}}.

\bibitem{Atalay2024}
S.~Atalay, E.~Schnaubelt, M.~Wozniak, J.~Dular, G.~Zachou, S.~Schöps, A.~Verweij, {An open-source 3D FE quench simulation tool for no-insulation HTS pancake coils}, Superconductor Science and Technology 37~(6) (2024) 065005.
\newblock \href {https://doi.org/10.1088/1361-6668/ad3f83} {\path{doi:10.1088/1361-6668/ad3f83}}.

\bibitem{wozniak2024}
M.~Wozniak, E.~Schnaubelt, J.~Dular, E.~Ravaioli, A.~Verweij, {Quench Co-Simulation of Canted Cos-Theta Magnets}, IEEE Transactions on Applied Superconductivity 34~(3) (2024) 1--5.
\newblock \href {https://doi.org/10.1109/TASC.2023.3338142} {\path{doi:10.1109/TASC.2023.3338142}}.

\bibitem{geuzaine2009gmsh}
C.~Geuzaine, J.-F. Remacle, {Gmsh}: A three-dimensional finite element mesh generator with built-in pre- and post-processing facilities, International Journal for Numerical Methods in Engineering 79~(11) (2009) 1309--1331.
\newblock \href {https://doi.org/10.1002/nme.2579} {\path{doi:10.1002/nme.2579}}.

\bibitem{getdp_1998}
P.~Dular, C.~Geuzaine, F.~Henrotte, W.~Legros, {A general environment for the treatment of discrete problems and its application to the finite element method}, IEEE Transactions on Magnetics 34~(5) (1998) 3395--3398.
\newblock \href {https://doi.org/10.1109/20.717799} {\path{doi:10.1109/20.717799}}.

\bibitem{GetDP}
P.~Dular, C.~Geuzaine, {GetDP}, \url{https://gitlab.onelab.info/getdp/getdp/-/tree/getdp_4_0_0_rc1} (2025).

\bibitem{schnaubelt_cerngetdp}
E.~Schnaubelt, \href{https://cern.ch/cerngetdp}{{CERNGetDP}}, \,Version: 2025.12.1.
\newline\urlprefix\url{https://cern.ch/cerngetdp}

\bibitem{Monk2003}
P.~Monk, {Finite Element Methods for Maxwell's Equations}, Oxford University Press, 2003.
\newblock \href {https://doi.org/10.1093/acprof:oso/9780198508885.001.0001} {\path{doi:10.1093/acprof:oso/9780198508885.001.0001}}.

\bibitem{Driesen2001}
J.~Driesen, R.~Belmans, K.~Hameyer, {Finite-element modeling of thermal contact resistances and insulation layers in electrical machines}, IEEE Transactions on Industry Applications 37~(1) (2001) 15--20.
\newblock \href {https://doi.org/10.1109/28.903121} {\path{doi:10.1109/28.903121}}.

\bibitem{Schnaubelt2025}
E.~Schnaubelt, A.~Vitrano, M.~Wozniak, E.~Ravaioli, A.~Verweij, S.~Schöps, {Transient finite element simulation of accelerator magnets using thermal thin shell approximation}, Superconductor Science and Technology 38~(6) (2025) 065003.
\newblock \href {https://doi.org/10.1088/1361-6668/add841} {\path{doi:10.1088/1361-6668/add841}}.

\bibitem{Schnaubelt2023a}
E.~Schnaubelt, M.~Wozniak, S.~Schöps, {Thermal thin shell approximation towards finite element quench simulation}, Superconductor Science and Technology 36~(4) (2023) 044004.
\newblock \href {https://doi.org/10.1088/1361-6668/acbeea} {\path{doi:10.1088/1361-6668/acbeea}}.

\bibitem{Chan2010}
W.~K. Chan, P.~J. Masson, C.~Luongo, J.~Schwartz, {Three-Dimensional Micrometer-Scale Modeling of Quenching in High-Aspect-Ratio $\hbox{YBa}_{2}\hbox{Cu}_{3}\hbox{O}_{7 - \delta}$ Coated Conductor Tapes—Part I: Model Development and Validation}, IEEE Transactions on Applied Superconductivity 20~(6) (2010) 2370--2380.
\newblock \href {https://doi.org/10.1109/TASC.2010.2072956} {\path{doi:10.1109/TASC.2010.2072956}}.

\bibitem{Schnaubelt2024}
E.~Schnaubelt, S.~Atalay, M.~Wozniak, J.~Dular, C.~Geuzaine, B.~Vanderheyden, N.~Marsic, A.~Verweij, S.~Schöps, {Magneto-Thermal Thin Shell Approximation for 3D Finite Element Analysis of No-Insulation Coils}, IEEE Transactions on Applied Superconductivity 34~(3) (2024) 1--6.
\newblock \href {https://doi.org/10.1109/TASC.2023.3340648} {\path{doi:10.1109/TASC.2023.3340648}}.

\bibitem{Foussat2026}
A.~Foussat, N.~Sala, S.~Gowrishankar, J.~C. Perez, A.~Haziot, C.~Fernandes, L.~Fiscarelli, M.~Wozniak, E.~Ravaioli, O.~Id’Bahmane, L.~Gentini, M.~Guinchard, O.~S. De~Frutosins, E.~F. Mora, E.~Todesco, S.~Farinon, A.~Pampaloni, M.~Sorbi, R.~U. Valente, {High Field Cos-Theta FalconD-C Dipole Magnet Development at CERN}, IEEE Transactions on Applied Superconductivity 36~(3) (2026) 1--6.
\newblock \href {https://doi.org/10.1109/TASC.2025.3628537} {\path{doi:10.1109/TASC.2025.3628537}}.

\bibitem{Bortot2017}
L.~Bortot, B.~Auchmann, I.~{Cortes Garcia}, A.~M. Fernandez~Navarro, M.~Maciejewski, M.~Prioli, S.~Schöps, A.~P. Verweij, {A 2-D Finite-Element Model for Electrothermal Transients in Accelerator Magnets}, IEEE Transactions on Magnetics 54~(3) (2018) 1--4.
\newblock \href {https://doi.org/10.1109/TMAG.2017.2748390} {\path{doi:10.1109/TMAG.2017.2748390}}.

\bibitem{Verweij2006}
A.~Verweij, {CUDI: A model for calculation of electrodynamic and thermal behaviour of superconducting Rutherford cables}, Cryogenics 46~(7) (2006) 619--626.
\newblock \href {https://doi.org/10.1016/j.cryogenics.2006.01.009} {\path{doi:10.1016/j.cryogenics.2006.01.009}}.

\bibitem{SchnaubeltPhD}
E.~M. Schnaubelt, {Models and Methods for Transient Magneto-Thermal Finite Element Simulation of Superconducting Magnets}, Doctoral dissertation, Technische Universität Darmstadt, Darmstadt, Germany (October 2025).
\newblock \href {https://doi.org/10.26083/tuprints-00031402} {\path{doi:10.26083/tuprints-00031402}}.

\bibitem{Chorowski2006}
M.~Chorowski, S.~Pietrowicz, R.~van Weelderen, Towards a better understanding of the physics of the two-volume model of accelerator magnet quench thermohydraulics, Cryogenics 46~(7) (2006) 581--588.
\newblock \href {https://doi.org/10.1016/j.cryogenics.2006.01.013} {\path{doi:10.1016/j.cryogenics.2006.01.013}}.

\bibitem{Bottura2024}
L.~Bottura, F.~Zimmermann, {High Energy LHC Machine Options in the LHC Tunnel}, 2024, Ch. Chapter 26, pp. 367--396.
\newblock \href {https://doi.org/10.1142/9789811280184_0026} {\path{doi:10.1142/9789811280184_0026}}.

\bibitem{Xu2025}
Z.~Xu, H.~Wang, Z.~Feng, H.~Wu, J.~Xiao, Q.~Chen, S.~Wang, J.~Cheng, L.~Wang, Y.~Wang, J.~Liu, C.~Xu, Q.~Wang, {Design and simulation of a 7.0T conduction cooled superconducting magnet}, Scientific Reports 15 (2025) 15699.
\newblock \href {https://doi.org/10.1038/s41598-025-00643-w} {\path{doi:10.1038/s41598-025-00643-w}}.

\bibitem{Shimizu2022}
H.~Shimizu, Y.~Arimoto, Z.~Zong, N.~Ohuchi, K.~Umemori, A.~Yamamoto, N.~Kimura, V.~Kashikhin, {Study on Conduction Cooling of Superconducting Magnets for the ILC Main Linac}, IEEE Transactions on Applied Superconductivity 32~(6) (2022) 1--5.
\newblock \href {https://doi.org/10.1109/TASC.2022.3155487} {\path{doi:10.1109/TASC.2022.3155487}}.

\bibitem{Nezuja2026}
H.~Nezuka, T.~Uto, H.~Takewa, T.~Shimonosono, A.~Badel, K.~Takahashi, T.~Okada, Y.~Tsuchiya, S.~Awaji, {Detailed Manufacturing Design of REBCO Insert and Cooling Performance of 33T Cryogen-Free Superconducting Magnet}, IEEE Transactions on Applied Superconductivity 36~(3) (2026) 1--5.
\newblock \href {https://doi.org/10.1109/TASC.2025.3641099} {\path{doi:10.1109/TASC.2025.3641099}}.

\bibitem{BorgesDeSousa2026}
P.~Borges~de Sousa, X.~Gallud, A.~Petrovic, L.~Delprat, B.~Bradu, R.~van Weelderen, L.~Tavian, D.~Delikaris, {Toward a Reduced Helium Content Cryogenic Cooling Scheme at 4.5 K for CERN’s FCC-hh Accelerator}, IEEE Transactions on Applied Superconductivity 36~(3) (2026) 1--7.
\newblock \href {https://doi.org/10.1109/TASC.2025.3604763} {\path{doi:10.1109/TASC.2025.3604763}}.

\bibitem{Todesco2018}
E.~Todesco, M.~Annarella, G.~Ambrosio, G.~Apollinari, A.~Ballarino, H.~Bajas, M.~Bajko, B.~Bordini, R.~Bossert, L.~Bottura, E.~Cavanna, D.~Cheng, G.~Chlachidze, G.~De~Rijk, J.~DiMarco, P.~Ferracin, J.~Fleiter, M.~Guinchard, A.~Hafalia, E.~Holik, S.~Izquierdo~Bermudez, F.~Lackner, M.~Marchevsky, C.~Loeffler, A.~Nobrega, J.~C. Perez, S.~Prestemon, E.~Ravaioli, L.~Rossi, G.~Sabbi, T.~Salmi, F.~Savary, J.~Schmalzle, S.~Stoynev, T.~Strauss, M.~Tartaglia, G.~Vallone, G.~Velev, P.~Wanderer, X.~Wang, G.~Willering, M.~Yu, Progress on hl-lhc nb3sn magnets, IEEE Transactions on Applied Superconductivity 28~(4) (2018) 1--9.
\newblock \href {https://doi.org/10.1109/TASC.2018.2830703} {\path{doi:10.1109/TASC.2018.2830703}}.

\bibitem{DeGersem2020}
H.~{De Gersem}, I.~{Cortes Garcia}, L.~A.~M. D'Angelo, S.~Schöps, {Magnetodynamic Finite-Element Simulation of Accelerator Magnets} (2020).
\newblock \href {https://doi.org/10.48550/arXiv.2006.10353} {\path{doi:10.48550/arXiv.2006.10353}}.

\bibitem{jackson1998}
J.~D. Jackson, {Classical Electrodynamics}, 3rd Edition, Wiley, New York, NY, 1998.

\bibitem{CAMPBELL1982}
A.~Campbell, {A general treatment of losses in multifilamentary superconductors}, Cryogenics 22~(1) (1982) 3--16.
\newblock \href {https://doi.org/10.1016/0011-2275(82)90015-7} {\path{doi:10.1016/0011-2275(82)90015-7}}.

\bibitem{Dular2025}
J.~Dular, A.~Verweij, M.~Wozniak, {Reduced order hysteretic magnetization model for composite superconductors}, Superconductor Science and Technology 38~(3) (2025) 035017.
\newblock \href {https://doi.org/10.1088/1361-6668/adb5cc} {\path{doi:10.1088/1361-6668/adb5cc}}.

\bibitem{Glock2025}
A.~Glock, J.~Dular, A.~Verweij, M.~Wozniak, {Reduced Order Hysteretic Flux Model for Transport Current Homogenization in Composite Superconductors}, IEEE Transactions on Magnetics (2025).
\newblock \href {https://doi.org/10.1109/TMAG.2025.3613877} {\path{doi:10.1109/TMAG.2025.3613877}}.

\bibitem{Dular2025b}
J.~Dular, A.~Glock, A.~Verweij, M.~Wozniak, {Distributed Inter-Strand Coupling Current Model for Finite Element Simulations of Rutherford Cables} (2025).
\newblock \href {http://arxiv.org/abs/2510.24618} {\path{arXiv:2510.24618}}, \href {https://doi.org/10.48550/arXiv.2510.24618} {\path{doi:10.48550/arXiv.2510.24618}}.

\bibitem{Parent2008}
G.~Parent, P.~Dular, J.-P. Ducreux, F.~Piriou, {Using a Galerkin Projection Method for Coupled Problems}, IEEE Transactions on Magnetics 44~(6) (2008) 830--833.
\newblock \href {https://doi.org/10.1109/TMAG.2008.915798} {\path{doi:10.1109/TMAG.2008.915798}}.

\bibitem{Hahn2024}
R.~Hahn, E.~Schnaubelt, M.~Wozniak, C.~Geuzaine, S.~Sch{\"o}ps, {Mortar Thin Shell Approximation for Analysis of Superconducting Accelerator Magnets} (2024).
\newblock \href {https://doi.org/10.48550/arXiv.2405.01076} {\path{doi:10.48550/arXiv.2405.01076}}.

\end{thebibliography}

\end{document}